\documentclass{elsart}

\usepackage{graphicx}

\usepackage{amssymb}

\begin{document}

\begin{frontmatter}



\title{Spatio-temporal Dynamics in the Origin of Genetic Information}


\author{Pan-Jun Kim}
\ead[url]{http://stat.kaist.ac.kr/\~{}pj/origin.html},
\author{Hawoong Jeong\corauthref{cor}}
\corauth[cor]{Corresponding author. Tel:+82-42-869-2543; fax:+82-42-869-2510.}
\ead{hjeong@kaist.ac.kr}

\address{Department of Physics, Korea Advanced Institute of Science and Technology,
Daejeon 305-701, South Korea}

\begin{abstract}

We study evolutionary processes induced by spatio-temporal dynamics in prebiotic evolution.
Using numerical simulations we demonstrate that hypercycles emerge from
complex interaction structures in multispecies systems. In this work
we also find that `hypercycle hybrid' protects the hypercycle from its environment
during the growth process.
There is little selective advantage for one hypercycle to maintain coexistence with others.
This brings the possibility of the outcompetition between hypercycles
resulting in the negative effect on information diversity. To enrich the information
in hypercycles, symbiosis with parasites is suggested. It is shown that symbiosis with
parasites can play an important role in the prebiotic immunology.	

\end{abstract}

\end{frontmatter}


\section{Introduction}

The appearance of a molecule which is capable of replicating itself is probably
the most fundamental event in the history of life.
One of the candidates for the prebiotic genes, RNA is both the carrier
of genetic information and the molecule with
biological activities of which variety becomes an interesting topic recently \cite{Guy}.
Eigen and coworkers were
the first to address the existence of the information threshold
in prebiotic evolution that the length of a
molecule (polynucleotide) is limited due to the finite replication accuracy per nucleotide
\cite{aEigen}.
The maximum length of the molecules attained by the process of Darwinian selection
seems to be too short for a genetic message to encode a functional protein.
In their hypercycle theory Eigen suggests that
if molecules catalyze the replication of each other in a cyclic way,
the information threshold can be crossed \cite{aEigen}.
No molecule in the hypercycle can outcompete another
because they are forced to cooperate. Each molecule is still
bound to the maximum string-length, but the molecules can combine their information and
thus the information threshold can be crossed.

However, there are two major problems with this idea. The first problem is that hypercycles with five or more species show a limit cycle behavior.
This implies that large hypercycles are unstable because some
species may become extinct. The second problem is that providing catalytic support
to other species is in fact an altruistic behavior, therefore, they are extremely vulnerable
to the presence of parasites which are species that do not reciprocate the catalytic support
they receive.

Boerlijst and Hogeweg \cite{Boerlijst} have shown that spatial self-structuring can solve
these extinction and parasite problems. They added the diffusive behavior of each
species to the early model of hypercycle. In this spatially diffusive model, hypercycles with
five or more species spontaneously generate spiral waves, rotating pattern of all species
in the hypercycle. With the help of this rotating spiral wave global
extinction of species no longer takes place.
Furthermore, it turns out that spiral waves are resistent to parasite invasions.
Because the molecules in the center of a spiral generate an offspring of the entire spiral
in radial direction,
it is difficult for a parasite to grow towards the center of the spiral.

Their pioneering work presents three important points. First, natural selection is
effectively driven by spatio-temporal dynamics
\cite{Boerlijst,bBoerlijst,Savill,Rauch}, not only by the
population-driven competition between species. For example, Boerlijst
and Hogeweg observed that increasing of diffusion rate makes the spiral patterns bigger,
enhancing the resistence to invasion of parasites. Second, as a consequence of
self-structuring, natural selection occurs at the level of the community but not
of the individual \cite{Johnson}. A community maintains some minimum level of
integrity for a long period enough for natural selection to act on. Integrity can be maintained
as a form of nonrandom pattern in the spatial arrangement of individuals.
Each spiral behaves like super-organism
whose boundary is determined by the molecules in the periphery. Therefore the evolutionary
attractors do not convey a fitness benefit to individual species but to the community
determined by the spiral
\cite{bBoerlijst,cBoerlijst}. Third, they showed spatio-temporal dynamics of interaction
networks can bring
out nontrivial behaviors. The hypercycle which consists
of a single cyclic interaction structure seems to be fatal to parasite invasions, but it
turns out that the hypercycle is very robust when we consider its
spatio-temporal dynamics. However, little is known about spatio-temporal dynamics
for the case of more complex interaction networks
in this respect, while many research groups have studied
several properties of complex networks
recently \cite{Reka,Manru}.

In this paper, we addressed several issues which should be resolved
to fully understand the spatio-temporal dynamics of multispecies interaction network.
Boerlijst and Hogeweg let parasite
invade hypercycles
after full development of their spatial structures. But if parasite-like species is
present before self-structuring of hypercycles, can hypercycles
successfully drive away their initially embedded parasites? Also, if
the multispecies system has the complex interaction structure (network) rather than a single
cyclic structure as in the hypercycle model, is it possible that some cyclic sub-structures
outcompete other sub-structures by the process of spiral-formation? We expect
that the complex interaction structure would be separated
into many cyclic sub-structures by self-structuring at first.
And then competition between these cyclic sub-structures would select sub-structure
with more efficient species. 
Our scenario on prebiotic evolution is different from that of Jain and Krishna
\cite{Jain,bJain} in which selection on species in the same giant structure containing
autocatalytic sets is suppressed during structure development. In their scenario, evolutionary
unstability is unavoidable by a collapse of the giant structure due to
the selection at the level
of individuals in the complex interaction network.

In the next section we investigate the competition between initially embedded parasites
and hypercycles and find the mechanism of
separation of cyclic sub-structures using a simple model structure.
Using numerical simulations, in Section 3 we demonstrate an
emergence of information communities in prebiotic evolution. And we find
that the `hypercycle hybrid' protects the hypercycle from its environment
during the development process.
In Section 4 by investigating interaction between developed communities
we find that
self-interest of each community can discourage inheritance of diverse information.
To recover information diversity, in Section 5 we introduce
symbiosis with parasites and show that these embeded parasites
can play an important role on the
immunology of the community. The final section is devoted to conclusions.

\section{Simple Trials}
To study the spatio-temporal dynamics of hypercycles we use the following model
equation \cite{dBoerlijst}.
\begin{displaymath}
\frac{d X_{i}}{d t}= -\delta_{i} X_{i}+ \big(1-\sum_{k=1}^{N}X_k \big)\big(\rho_{i}+
\sum_{j}\kappa _{i,j}X_{j}\big)X_{i} +D_{i}\nabla^{2}X_{i}\,\, .
\end{displaymath}
We assume there are $N$ different types
of species and $X_{i}$ denotes the concentration of species $i$ at a given site.
$\delta_{i}$ stands for the spontaneous decay rate of species $i$, $\rho_{i}$ for the self-replication
rate, and $D_{i}$ for the diffusion coefficient.
$\kappa_{i,j}$
is the rate of replication of species $i$ catalyzed by species $j$.
The first (second) term of rhs corresponds to the decay (growth) rate of population of
species.
The term $\big(1-\sum_{k=1}^{N}X_k\big)$ limits the population of species at the same site
due to finite resources. The final term shows diffusive behavior of species $i$
which is responsible for spatial pattern formation.
For numerical simulations we use a rectangular grid of $147 \times 147$ sites with
Neumann boundary conditons. And 
we assume that species $i$ is extinct when $X_{i}$ is small enough (e.g., $X_{i}<10^{-3}$).
We use
the integration time step $D_{max} \Delta t/(\Delta x)^{2}=0.1$ where we choose $D_{max}$ as the maximum
value of $D_{i}$ ( $i=1,2,\dots,N$ ), $\Delta x =1$. We use the parameters
$\delta_{i}=1$, $\rho_{i}=2$, $D_{i}=1$,
$\kappa_{i,j}=0$ or $500$ (depending on the existence of catalytic support
to species $i$ from species $j$) unless specified \cite{com}.

\begin{figure}[h]
\begin{center}
\includegraphics[height=0.5\textwidth]{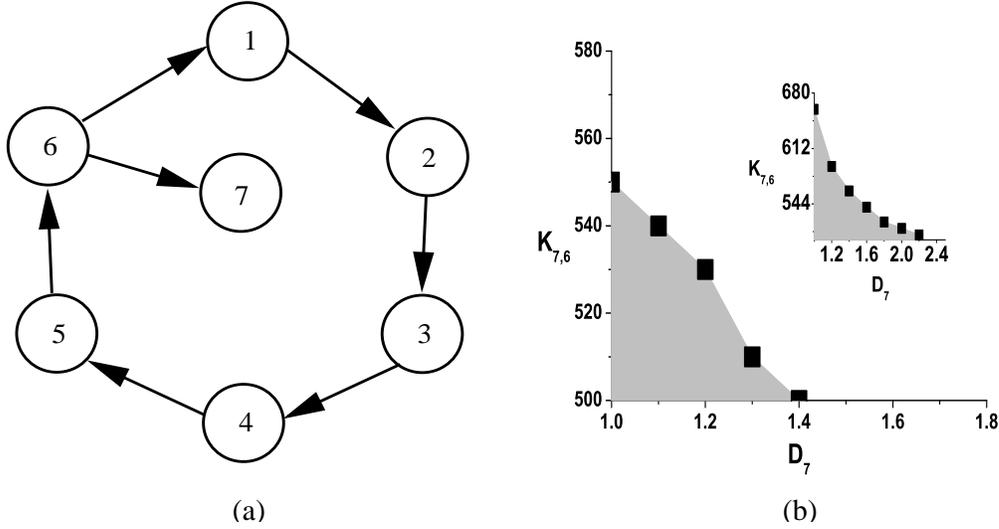}
\caption{(a) Schematic diagram of a hypercycle with the parasite-like species 7.
(b) Phase diagram for the case A (inset for the case B). The horizontal axis for $D_{7}$,
the vertical axis for $\kappa_{7,6}$. Shadowed area below the filled
squares corresponds to the phase where
parasite species is outcompeted. Refer to the main text for the case A and B.}
\label{1}
\end{center}
\end{figure}

One necessary condition for cyclic sub-structures being favored in complex interaction
structure would be the stability of the cyclic structures to initially embedded
parasites. To investigate this stability we consider the structure in Fig. \ref{1}(a).
Each arrow indicates
the direction of catalysis, e.g. $2 \rightarrow 3$ means that species 2 catalyzes
the replication of species 3 and, therefore concentration of species 3 increases.
Species 7 is a parasite because it does not
reciprocate the catalytic support given by species 6.
We study the following two cases. First, to test whether hypercycles can drive away
initially embedded parasites, we start with randomly assigned initial concentration containing
molecules of all species (case A). Second, as a comparison with the situation where
parasites invade
hypercycles after
fully development of spatial structure, we begin with randomly assigned initial concentration
of six species while parasite concentration remains zero.
And after spirals are fully developed, species in a half of the sites
are replaced by parasite species (case B) \cite{rnd}.

\begin{figure}[h]
\begin{center}
\includegraphics[height=0.33\textwidth]{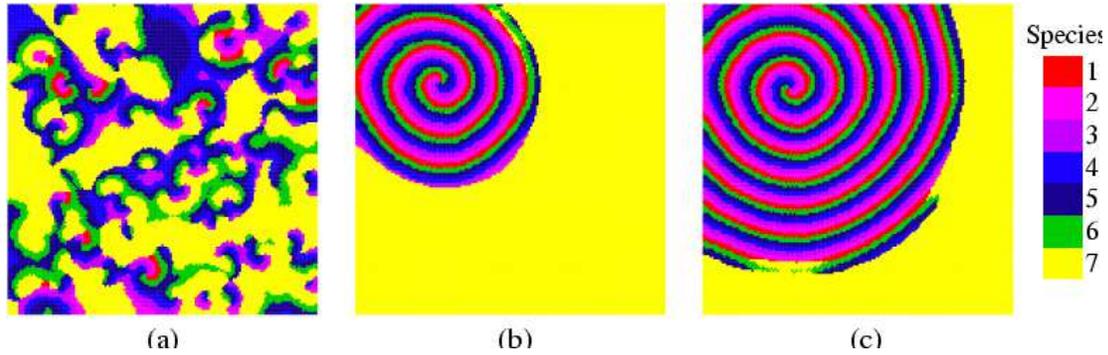}
\caption{Development of a spiral wave pattern with $D_{7}=1$ and $\kappa_{7,6}=530$
in the case A.
Painted with the color
corresponding to the species which has the larger population size than any other
species at
a given site. Following pictures for spatial patterns are painted by the same method.
(a) $t=10$ (b) $t=100$ (c) $t=200$.}
\label{2}
\end{center}
\end{figure}

By numerical simulations it is identified that initially embedded parasites are also effectively
outcompeted although it is only possible in
the narrow parameter region when compared with case B
(see Fig. \ref{1}(b)).
It turns out that even if parasites are initially embedded,
spiral cores avoiding infection of parasites
in the early stage can drive away the parasites by radial growth
of the spiral waves (see Figs. \ref{2}(a--c)).

\begin{figure}[h]
\begin{center}
\includegraphics[height=0.47\textwidth]{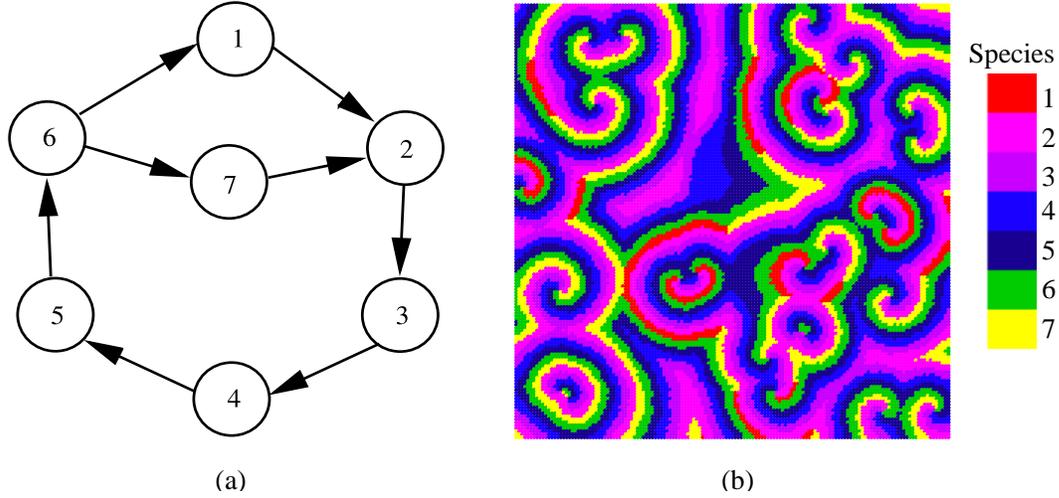}
\caption{(a) Schematic diagram of simply entangled hypercycles.
(b) Development of two different spirals from entangled hypercycle network.}
\label{3}
\end{center}
\end{figure}

\begin{figure}[h]
\begin{center}
\includegraphics[height=0.33\textwidth]{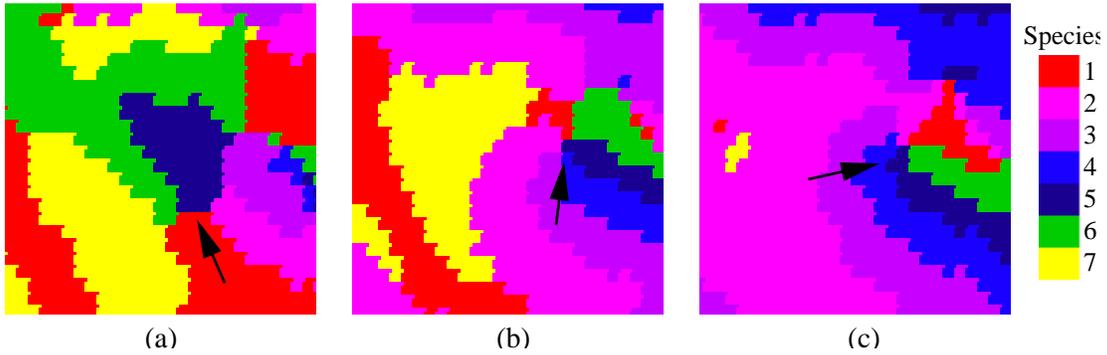}
\caption{(a) Competition between two cyclic sub-structures.
Magnified pattern near the spiral core, at $t=3$. The arrow in the figure indicates
the spiral core with $1 \rightarrow 2 \rightarrow 3 \rightarrow 4 \rightarrow 5 \rightarrow 6 \rightarrow 1$.
Species 7 is slightly out of the core. (b) At $t=5$ species 7 cannot invade the spiral core.
(c) At $t=6$ species 7 is washed out by the environment.}
\label{4}
\end{center}
\end{figure}

Next, we consider the structure in Fig. \ref{3}(a) where species 7 is equivalent to species
1 in their parameters.
Separation of the cyclic
sub-structures 
is successful by the spiral formation from molecules with randomly assigned concentration
(see Fig. \ref{3}(b)). There are two kinds of spirals. One is composed of $1 \rightarrow 2 \rightarrow 3 \rightarrow 4 \rightarrow 5 \rightarrow 6 \rightarrow 1$
and the other is composed of $7 \rightarrow 2 \rightarrow 3 \rightarrow 4 \rightarrow 5 \rightarrow 6 \rightarrow 7$.
Species 1 and 7 cannot coexist in the same spiral because
the species (either 1 or 7) slightly out of the spiral core
fails to invade the spiral core and is washed out from its environment
(see Figs. \ref{4}(a--c)). As a result species 1 and 7 cannot
coexist in the same spiral except for the case
where two kinds of spirals come in contact
with each other at peripheries. But even in that case separation
at the level of spiral cores is clear (see Fig. \ref{3}(b)) and separation of cyclic
sub-structures is completed.

In this section we examined the stability of the spiral against parasites in early stage,
and explored the possibility that the specific cyclic sub-structures
can be favored in the complex interaction networks.

\section{Emergence of Communities}

\begin{figure}[h]
\begin{center}
\includegraphics[height=1.1\textwidth]{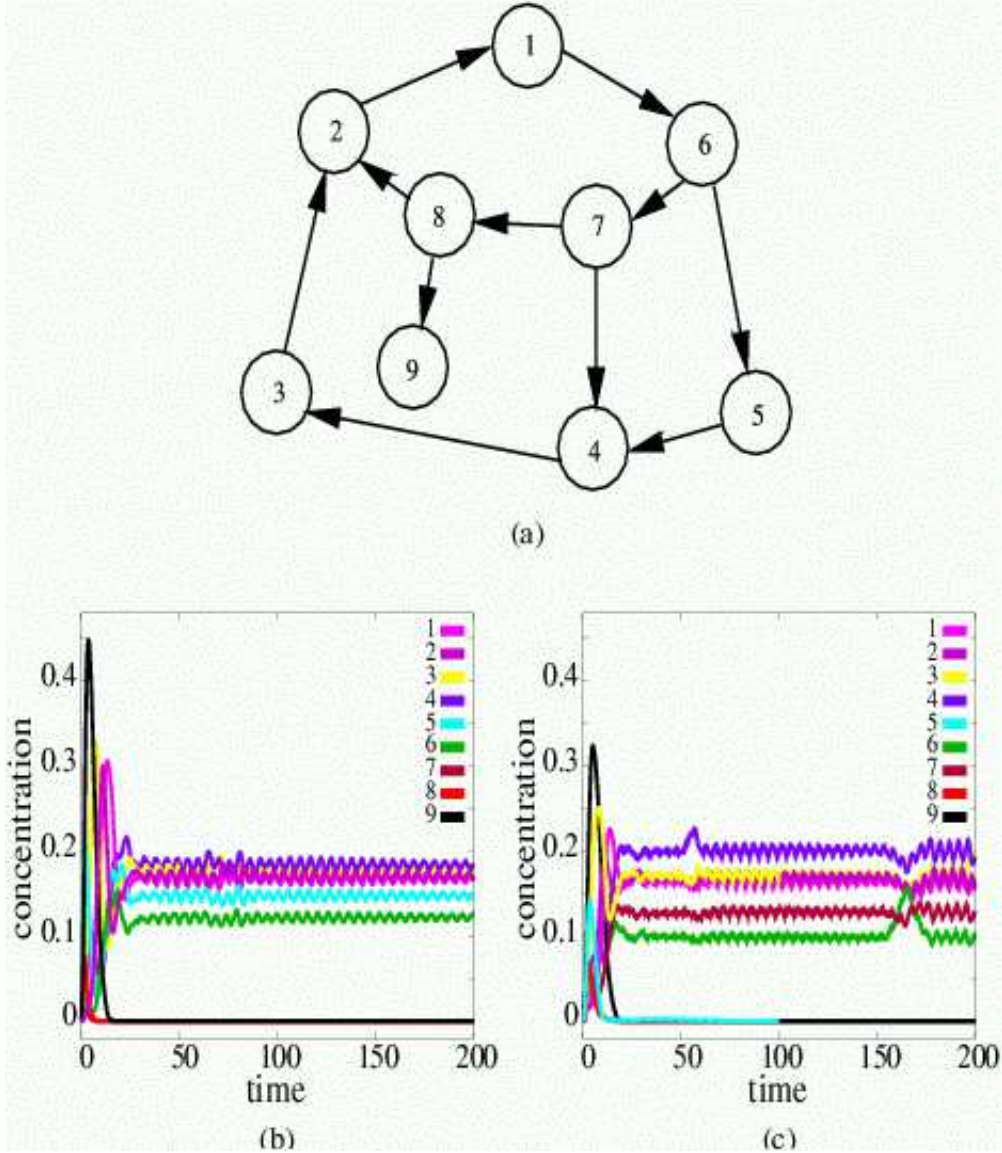}
\caption{(a) Schematic diagram of the model structure.
Population of each species after local changes, where the
horizontal axis is for
evaluation time, the vertical axis for concentration of species
averaged over the total sites.
(b) When $D_{5}=2$ the species in $1 \rightarrow 6 \rightarrow 5 \rightarrow 4 \rightarrow 3 \rightarrow 2 \rightarrow 1$
are populated while the rest of species become suppressed.
(c) When $\kappa_{7,6}=\kappa_{4,7}=1000$ the species in $1 \rightarrow 6 \rightarrow 7 \rightarrow 4 \rightarrow 3 \rightarrow 2 \rightarrow 1$
are populated while the rest of species become suppressed.}
\label{5}
\end{center}
\end{figure}

Next, we consider more complex model structure as in Fig. \ref{5}(a). There are three cyclic sub-structures,
$1 \rightarrow 6 \rightarrow 5 \rightarrow 4 \rightarrow 3 \rightarrow 2 \rightarrow 1$,
$1 \rightarrow 6 \rightarrow 7 \rightarrow 4 \rightarrow 3 \rightarrow 2 \rightarrow 1$ and
$1 \rightarrow 6 \rightarrow 7 \rightarrow 8 \rightarrow 2 \rightarrow 1$.
As expected in the previous section the cyclic
sub-structure repels other sub-structures and each cyclic sub-structure coexists in the
form of spirals. Each spiral can be considered as community where species in the same cycle
mainly interacts each other.

However, it is unlikely that every parameters of each species will
have the same value as system gets complex. We demonstrate that
selection process indeed acts on these cyclic sub-structures due to species dependent
parameters.
If we increase the diffusion coefficient of species 5 by $D_{5}=2$, it is observed
that $1 \rightarrow 6 \rightarrow 5 \rightarrow 4 \rightarrow 3 \rightarrow 2 \rightarrow 1$
outcompetes other cyclic sub-structures
(see Fig. \ref{5}(b)) \cite{sub}. If we increase the catalytic support from species 6 to species 7
and
from species 7 to species 4 by $\kappa_{7,6}=\kappa_{4,7}=1000$,
$1 \rightarrow 6 \rightarrow 7 \rightarrow 4 \rightarrow 3 \rightarrow 2 \rightarrow 1$
outcompetes other cyclic sub-structures (see Fig. \ref{5}(c)). As you can see,
small variation in local parameter can change whole dynamics.

We note that selection occurs at the community level but not at the individual level
because the individual-level selection is overridden by community
\cite{Johnson}. In the situation of Fig. \ref{5}(c), the species 6 gains little benefit
in
$1 \rightarrow 6 \rightarrow 7 \rightarrow 4 \rightarrow 3 \rightarrow 2 \rightarrow 1$
cycle because of its altruistic behavior in this community
(see the population of species 6 in Fig. \ref{5}(c)). But since the species 6 enhances the fitness
of the entire community,
the species 6 involved in
$1 \rightarrow 6 \rightarrow 7 \rightarrow 4 \rightarrow 3 \rightarrow 2 \rightarrow 1$
is selected in the final stage rather than the species 6 in
$1 \rightarrow 6 \rightarrow 5 \rightarrow 4 \rightarrow 3 \rightarrow 2 \rightarrow 1$.

In the above example, we observed competition between cyclic sub-structures and
resulting selection process. From these observations we find that
we don't need to affect overall properties of the entire community to select a particular
community.
We just need to adjust local properties
of the interaction, e.g. diffusion coefficient of one node (Fig. \ref{5}(b))
or catalytic support across a few links (Fig. \ref{5}(c)). This `local
strategy' has the important meaning for the control scheme on networks (control process has
been an attractive topic to many research groups
\cite{Ditto,delay1,pattern1,pattern4},
but controlling networks is not such an explored field yet) and reflects the
self-structuring nature of the autocatalytic network.

\begin{figure}[h]
\begin{center}
\includegraphics[height=0.9\textwidth]{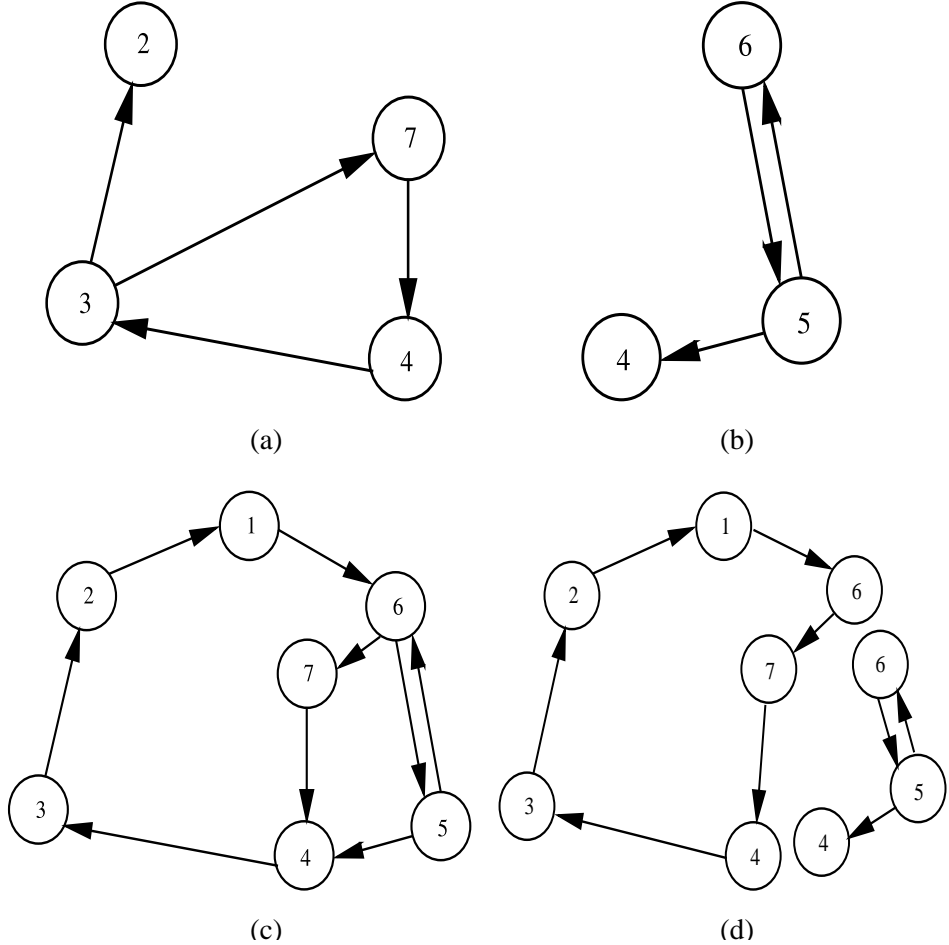}
\caption{(a) The fittest structure when we assign the catalytic support from species 3 to
species 7 in Fig. \ref{5}(a).
(b--c) Emergent sub-structures when assigning the catalytic support from species 5 to
species 6 in Fig. \ref{5}(a).
(d) The composite structure in (c) consists of two sub-structures.
Community with $6 \rightleftharpoons 5 \rightarrow 4$ is buried in the spiral formed by
$1 \rightarrow 6 \rightarrow 7 \rightarrow 4 \rightarrow 3 \rightarrow 2 \rightarrow 1$.}
\label{6}
\end{center}
\end{figure}

However, the above work does not guarantee that single-cyclic sub-structure is
always selected as a final fittest community. In fact
we found that spiral pattern is unstable under the presence of the flat generated by
a short hypercycle of 2 or 3 species (flat is stationary spatial-structure
dominated by few, less than 3 species).
For example, if we assign the
catalytic support from species 3 to species 7 in Fig. \ref{5}(a) the simple
structure in Fig. \ref{6}(a)
outcompetes other sub-structures in Fig. \ref{5}(a). The parasite-like species 2 in Fig. \ref{6}(a)
survives because the flat generated by a short hypercycle cannot effectively drive away
its parasites contrary to the spiral generated by a long hypercycle. This is consistent
with the observation of Boerlijst and Hogeweg that a short cycle with dangled parasites
frequently emerges from randomly connected structures \cite{cBoerlijst}.

The fact that the flat generated by 2 or 3 species outcompetes the spiral generated
by more species conflicts with the information-threshold crossing
introduced in Section 1. It is unnatural that
short cycles are the fittests in prebiotic evolution.

Let us consider the structure assigning the catalytic support from species 5
to species 6 in the structure of Fig. \ref{5}(a). Then two sub-structures are emerging; one is
the short cycle with a parasite as expected (see Fig. \ref{6}(b)), and the other is the
composite structure (see Fig. \ref{6}(c)) which can be separated into
two structures as in Fig. \ref{6}(d).

\begin{figure}[h]
\begin{center}
\includegraphics[height=0.47\textwidth]{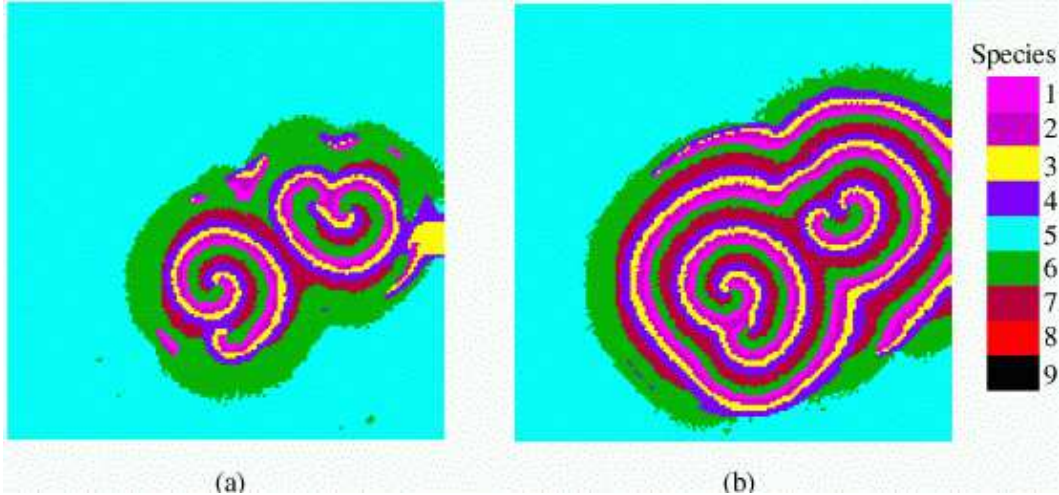}
\caption{Hypercycle hybrid between spiral from the structure in Fig. \ref{6}(c)
and flat from the short cycle in Fig. \ref{6}(b) at
(a) $t=100$ and (b) $t=200$. At the final stage the flat is outcompeted by the spiral completely.}
\label{7}
\end{center}
\end{figure}

We find that species 5 in the structure of Fig. \ref{6}(c) is eventually eliminated, and the flat formed by the short cycle in Fig. \ref{6}(b) is outcompeted by the spiral in Fig. \ref{6}(c) (see Figs. \ref{7}(a) and (b)).
At the initial stage, the flat is rapidly expanded and surrounds the spiral. The flat protects the spiral from the outside environment, otherwise
the environment may intimidate the spiral to be extinct. This `hypercycle hybrid'
between spiral and flat state is
maintained until the spiral becomes fully developed and then the flat is
outcompeted by the spiral.

Here we present how the hypercycle hybrid emerges from the complex structure. A short cycle cannot
suppress its parasites by the mechanism of spiral-formation. These dangling parasites weaken
the activity of the short cycle, thus the short cycle cannot outcompete the long hypercycle containing these parasites. For example, species 4 and 7 in Fig. \ref{6}(c) are not driven away
by the cycle
composed of species 5 and 6, so the hypercycle hybrid can exist as we described above.

As a consequence a hypercycle hybrid emerges naturally by the
short cycle with its parasites, and the hypercycle hybrid provides the selective advantage
to the long hypercycle by protecting its early growth process from the environmental species.
Selection process on hypercycle hybrid is also expected as previously demonstrated in the case of pure hypercycles in Fig. \ref{5}(b) and (c).

\section{Maintenance of Communities}

\begin{figure}[h]
\begin{center}
\includegraphics[height=0.33\textwidth]{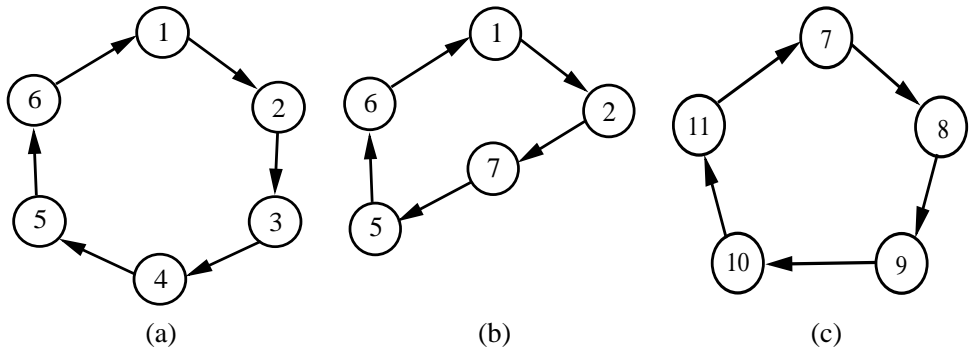}
\caption{Interrelation between communities toward different
dynamical consequences. Cycle (a) and (b) are interrelated via species
$1$, $2$, $5$, $6$.
In contrast, cycle (a) and (c) are completely disjoint.}
\label{add}
\end{center}
\end{figure}

After communities are developed completely, interaction between communities becomes an important factor to determine the evolutionary direction of the communities.
Let us consider interaction between two emergent hypercycles;
cycle A ($1 \rightarrow 2 \rightarrow 3 \rightarrow 4 \rightarrow 5 \rightarrow 6 \rightarrow 1$)
and cycle B ($1 \rightarrow 2 \rightarrow 7 \rightarrow 5 \rightarrow 6 \rightarrow 1$)
in Figs. \ref{add}(a) and (b).
We find that
cycle B outcompetes cycle A in the long time limit. One difference between these two cycles is
the number of species in the cycles, however we found that length of the cycle is not
responsible for this outcompetition.
For example, if we consider another cycle C ($7 \rightarrow 8 \rightarrow 9 \rightarrow 10 \rightarrow 11 \rightarrow 7$)
in Fig. \ref{add}(c) instead of cycle B,
it turns out that they cannot outcompete each other even though length of cycle C is shorter
than that of cycle A.
The important difference between these two cases is that
cycle A and C are disjoint while cycle A and B are interrelated via several species
(in this case, species 1, 2, 5, 6).
It turns out that the reason why cycle B outcompetes cycle A is related
with this coupling between two hypercycles.
More specifically we observe that population of species 4 is markedly decreased
when two kinds of spirals come in contact with each other;
at the boundary between two spirals,
species 5 bred by the short catalysis path across the species 7 (cycle B) suppresses
the population of species 4, which brings the elimination of cycle A.

The above observation can be interpreted as follows.
If there is no shared species/information between two communities (like cycle A and C),
nature prefers to conserve information by keeping two communities alive together. However,
if there is shared information (like cycle A and B),
nature prefers to select specific information rather than conserve information.

From the case between cycle A and C,
we notice that conservation of information is obtained by the
equivalent competition between spirals rather than by cooperation between spirals.
The community based on a spiral is maintained by dynamic process at its core, thus cooperation
at the spiral peripheries does not provide any qualitative changes to this system.
Therefore a spiral has little selective advantage when maintaining coexistence
with other spirals. This property imposes the possibility of the outcompetition between spirals
causing a negative effect on information diversity.

There are several ways to diversify genetic information. First, we can
allow an evolutionary change of species' character under the competitive ecological condition.
Imposed mutation at the spiral core can be effective in this respect.
In the spiral core, replication of molecules
occurs in an active manner expected to cause a number
of mutations,
and mutants in the spiral core are not easily
outcompeted than mutants in the spiral peripheries \cite{bBoerlijst}.
Second, we can compartmentalize the information, which was originally proposed to solve
the parasite problem in the hypercycle theory \cite{aEigen,Mikael}. Boerlijst and Hogeweg also
showed that compartmentation obtained by spatial gradient of molecular decay rate
increases the capacity for
information accumulation \cite{dBoerlijst}.
Each spiral in its own compartment has to overcome the surrounding barriers
in order to compete with the spirals in other compartments,
thus resulting in the suppressed competition.
Third, we can try endosymbiosis of information.
Endosymbiosis means that a symbiont dwells within the body of its symbiotic partner.
For the best known example, Margulis noted that the main internal structures of cells such as
mitochondria did not originate inside the cell, but reflect an endosymbiotic coupling
\cite{Goerner}. She showed that the basic path of biological evolution
is through the symbiosis of independent forms into more efficient and
more adaptive cooperatives.
We investigate the case by endosymbiosis of information in the next section.

\section{Innovation of Communities}

As presented in Section 4 spirals showing selfish behavior discourage inheritance of
diverse information. Here we investigate one possible solution to increase information diversity in the hypercycle
systems.
According to Section 2, when a parasite invades spirals the result is either of
the followings. The parasite outcompetes
spirals or the spirals drive away the parasite depending on choice of parameters.
What happens if the parasites which were considered to be toxic to
the spirals can catalyze i.e. give some benefit to some species of them?

\begin{figure}[h]
\begin{center}
\includegraphics[height=1\textwidth]{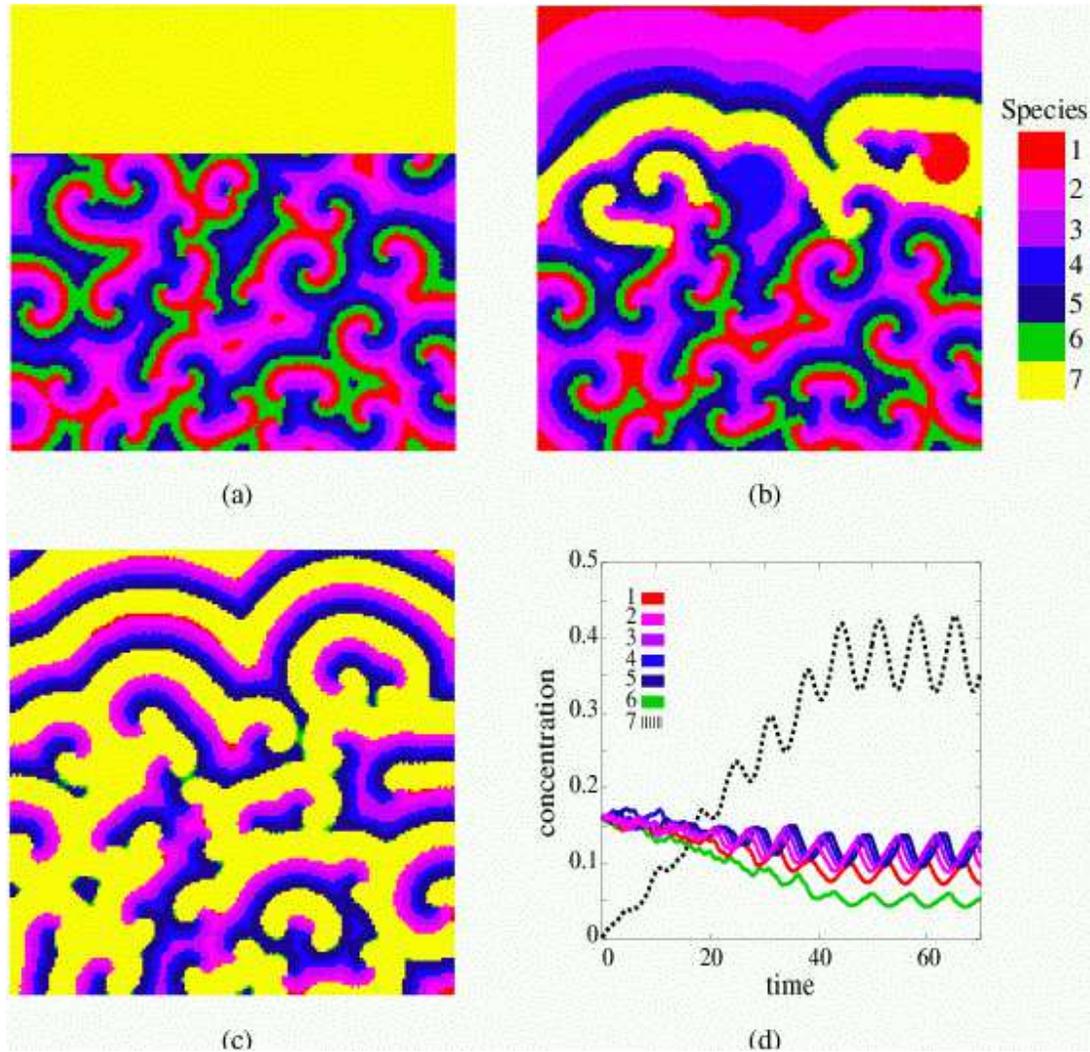}
\caption{When a parasite gives weak catalytic support to the hypercycle, endosymbiosis
between the hypercycle and its parasite is obtained.
Parameter $D_{7}=2$, $\kappa _{7,6}=1000$, $\kappa _{1,7}=100$ are used.
(a) At $t=0$, invasion of parasite was set from the top of developed spirals \cite{rnd}.
(b) At $t=10$, horizontal transfer of parasites. (c) At $t=100$, endosymbiosis between the
hypercycles and their parasites.
(d) Concentration of each species as a function of time. The horizontal axis is
for evaluation time,
the vertical axis for concentration of species averaged over the area occupied by the spirals.}
\label{8}
\end{center}
\end{figure}

Given the structure in Fig. \ref{1}(a),
we increase the diffusion coefficient and rate of catalyzed replication
of species 7 high enough to invade the spiral easily. Due to the
enhancement the species 7 will outcompete
the spirals. However, if we introduce relatively weak catalytic support
from species 7 to species 1, symbiosis between the spirals and
their parasites is acquired (see Figs. \ref{8}(a--c)).

The endosymbiosis over the communities is obtained by the horizontal
transfer of parasites. The retrovirus and transposon are examples of the movable genes
which are important in the history of evolution \cite{Stryer}. In this respect,
one can suspect that it is necessary for the
movable genes to show the
`parasitic' behavior. If we increase the catalytic support from species
7 to species 1 by $\kappa_{1,7}=500$, the species 7 fails to invade the spirals.
Therefore, the parasitic behavior is necessary for the movable genes to be in
symbiosis with spirals.

\begin{figure}[h]
\begin{center}
\includegraphics[height=0.47\textwidth]{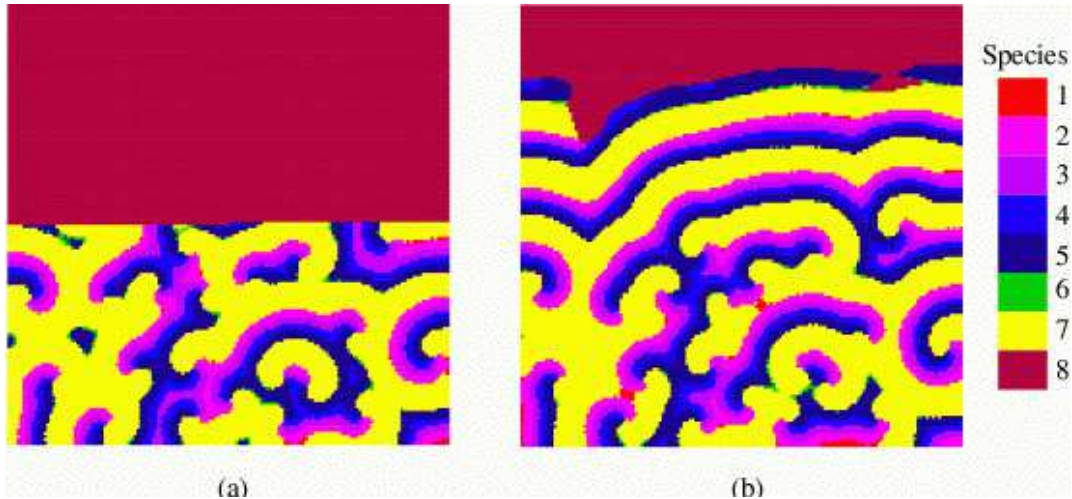}
\caption{Demonstration of vaccine effect (refer to the text).
(a) Invasion of species 8 from the top of Fig. \ref{8}(c) at $t=0$ \cite{rnd}.
(b) At $t=50$, the symbiotic unions are
driving species 8 away. At the final stage species 8 is completely outcompeted.}
\label{9}
\end{center}
\end{figure}

However, because of the parasitic behavior of species 7, the original species of spirals
are having difficulties such
that the populaton becomes depressed (see Fig. \ref{8}(d)). Nevertheless, there is a
benefit given to the spirals in the symbiosis with species 7. That is, let
us consider another species 8 which has the same properties as the species 7
except that it does not catalyze any species
in the spirals. Invasion of species 8 would be very
toxic to the spirals without species 7, but with the help of the symbiosis with species 7
the spirals can drive away
species 8 (see Figs. \ref{9}(a) and (b)), i.e. species 7 acts like `vaccine' against
the invasion of species 8.
This effect originates from the competition
between species 7 and species 8.
We suppose that this vaccine effect might be responsible for the
prebiotic immunology that should be much simpler than immunology of the present day
\cite{Perelson}.

\section{Conclusion}

We have studied spatio-temporal dynamics on the prebiotic evolution of genetic
information. In this paper we have investigated emergence, maintenance,
and innovation of the information communities. We recognize
that symbiosis is important on selection process in prebiotic evolution;
the short cycle which consists of only few symbionts fails to organize resistant structure
against parasites invasion. However, this helps the emergence of hypercycle hybrid providing
the selective advantage for a long hypercycle.
Endosymbiosis between
a hypercycle and its parasite shows interesting vaccine effect.
It is found that this symbiotic union is robust against the invasion of pure parasite.

Self-structuring of species by the complex interaction network leads to the
separation of cyclic sub-structures.
To select a particular sub-structure we only need to control local properties of the network
rather than overall properties of the entire specific sub-structure.
The integrity of each sub-structure reflects the emergent property
from the collective autocatalytic individuals.

The essential feature of autocatalysis is independent of its precise biochemical definition.
Therefore, study on autocatalysis would also be applicable to several area including
ecosystems, immune systems,
and social networks. We also want to emphasize that the role of self-structuring is not
restricted to the specific field --
prebiotic evolution. Self-structuring shows various phenomena
unexpected by our intuition based on simple ordinary differential equations.
In fact, rich nontrivial
results are reported
\cite{bBoerlijst,Savill,Rauch,Johnson,Hassell,Rohani,Kerr} in theoretical ecology.
Therefore we believe that our work can be of interest in many fields as well.

\section*{Acknowledgements}

We thank Tae-Wook Ko for useful comments on the manuscript.
The research was supported by the Ministry of Science and Technology through Korean Systems Biology Research Grant (M10309020000-03B5002-00000), and by KOSEF through Grant No. R08-2003-000-10285-0.





\begin{thebibliography}{00}
\bibitem{com}The specific parameters were selected because
they belong to the parameter region in which hypercycles can
be properly developed and
outcompete their initially embedded parasites. See Section 2 for more detail.
\bibitem{rnd}Instead of replacing the existing species by the invading species,
we also tried the randomly distributed invading species at that place to imitate more realistic
situation. The identical results are reproduced, probably by the partial differential equation
nature which promotes the discretized concentration to be flattened.
Somewhat different results may be possible in the cellular automata.
\bibitem{sub}The increased diffusion coefficient of species 5 promotes the transmission and
catalytic coupling of species 5 to other species in the early stage,
therefore bringing the selective advantage to the spiral development with species 5.
\end{thebibliography}
\end{document}